\documentclass[twocolumn]{aastex63}
\usepackage{graphicx}
\usepackage{latexsym}
\usepackage{amsfonts,amsmath,amssymb}
\usepackage{epstopdf}
\usepackage{natbib}
\usepackage{mathrsfs}
\usepackage{algorithm}
\usepackage[utf8]{inputenc}
\usepackage{hyperref}
\usepackage[noend]{algpseudocode}
\usepackage{xspace}

\accepted{June 4, 2020}
\submitjournal{ApJ}

\shorttitle{\Alu PSD Enrichment}
\shortauthors{Gaches et al.}

\newcommand{\Alu}{$^{26}$Al\xspace}
\newcommand{\Als}{$^{27}$Al\xspace}

\begin{document}

\title{Aluminum-26 Enrichment in the Surface of Protostellar Disks Due to Protostellar Cosmic Rays}

\correspondingauthor{Brandt A. L. Gaches}
\email{gaches@ph1.uni-koeln.de}

\author[0000-0003-4224-6829]{Brandt A. L. Gaches}
\affiliation{I. Physikalisches Institut, Universit\"{a}t zu K\"{o}ln, Z\"{u}lpicher Stra{\ss}e 77, 50937, K\"{o}ln, Germany}
\affiliation{Center of Planetary Systems Habitability, The University of Texas at Austin,  USA}

\author[0000-0001-6941-7638]{Stefanie Walch}
\affiliation{I. Physikalisches Institut, Universit\"{a}t zu K\"{o}ln, Z\"{u}lpicher Stra{\ss}e 77, 50937, K\"{o}ln, Germany}
\affiliation{Center for Data and Simulation Science, Universit\"{a}t zu K\"{o}ln, K\"{o}ln, Germany}

\author[0000-0003-1252-9916]{Stella S. R. Offner}
\affiliation{Department of Astronomy, The University of Texas at Austin, 2515 Speedway, Austin, Texas, 78712, USA}
\affiliation{Center of Planetary Systems Habitability, The University of Texas at Austin,  USA}

\author{Carsten M\"{u}nker}
\affiliation{Institut f\"{u}r Geologie und Mineralogie, Universit\"{a}t zu K\"{o}ln, Z\"{u}lpicher Stra{\ss}e 49b, 50674, K\"{o}ln, Germany}

\begin{abstract}
The radioactive decay of aluminum-26 (\Alu) is an important heating source in early planet formation. Since its discovery, there have been several mechanisms proposed to introduce \Alu into protoplanetary disks, primarily through contamination by external sources. We propose a local mechanism to enrich protostellar disks with \Alu through irradiation of the protostellar disk surface by cosmic rays accelerated in the protostellar accretion shock. We calculate the \Alu enrichment, [\Alu/\Als], at the surface of the protostellar disk in the inner AU throughout the evolution of low-mass stars, from M-dwarfs to proto-Suns. Assuming constant mass accretion rates, $\dot{m}$, we find that irradiation by MeV cosmic rays can provide significant enrichment on the disk surface if the cosmic rays are not completely coupled to the gas in the accretion flow. Importantly, we find that low accretion rates, $\dot{m} < 10^{-7}$ M$_{\odot}$ yr$^{-1}$, are able to produce canonical amounts of \Alu, $[^{26}{\rm Al}/^{27}{\rm Al}] \approx 5\times10^{-5}$. These accretion rates are experienced at the transition from Class I- to Class II-type protostars, when it is assumed that calcium-aluminum-rich inclusions condense in the inner disk. We conclude that irradiation of the inner disk surface by cosmic ray protons accelerated in accretion shocks at the protostellar surface may be an important mechanism to produce \Alu. Our models show protostellar cosmic rays may be a viable model to explain the enrichment of \Alu found in the Solar System.
\end{abstract}

\keywords{stars: protostars --- ISM: abundances --- (ISM:) cosmic rays --- meteorites, meteors, meteoroids --- protoplanetary disks}

\section{Introduction}
Short lived radioactive isotopes (SLRs) are radionuclides whose half-life is below 500 Myr, and any initial amounts in the solar system have since decayed below direct detection. Of these, aluminum-26 (\Alu), with a half-life of 0.717 Myr is of particular importance. \Alu within planetesimals can provide enough heat through radioactive decay to produce differentiated bodies with metal cores and silicate mantles \citep{wadhwa2007, dauphas2011}. %On Earth, the molten outer core helps to generate magnetic fields through a dynamo, and provides geothermal heat, a potential key ingredient for life. 
\Alu was first measured in the Allende meteorite calcium-aluminum-rich inclusions (CAIs) with the, now canonical, initial abundance ratio $[^{26}{\rm Al}/^{27}{\rm Al}] = {\rm n(^{26}Al)/n(^{27}Al)} \approx 5 \times 10^{-5}$ \citep{lee1976}. The \Alu/\Als abundance ratio in meteorites and CAIs is measured to be fairly consistent with the canonical value \citep{jacobsen2008}, although observations of chondrules show reductions of up to an order of magnitude \citep{tachibana2006, mishra2010, larsen2011}. The variable measured abundances of \Alu in chondrules at their time of formation are either ascribed to their prolonged formation interval \citep[e.g.][]{villeneuve2009} or, alternatively, to variable abundances of 26Al in different regions of the solar system \citep[e.g.][]{larsen2011}. The canonical ratio observed in meteorites shows an enhancement over the average \Alu/\Als ratio in the Galaxy \citep{diehl2006}. The complication is that, due to its short half-life, \Alu needs to be formed or mixed into the proto-Sun's protoplanetary disk shortly before the condensation the first solid matter in the solar system, as probed by CAIs.

Since it's discovery in meteorites by \citet{lee1976}, there has been much work to find a plausible enrichment mechanism. Such a mechanism should be able to explain the \Alu/\Als ratio and the fact that the ratio is found to be fairly homogeneous throughout the solar system \citep{dauphas2011}, although the homogeneity is debated \citep[see e.g.][]{larsen2011, krot2012}. Soon after the discovery, \citet{cameron1977} proposed that the enriched aluminium may come from the gas of a nearby supernova remnant. Enrichment via a nearby Type II supernova is an enticing mechanism due to the creation of both \Alu and $^{60}$Fe in significant quantities. Also, the appropriate aluminum isotope ratios could be produced in principle. Similarly, stellar winds have been proposed as an external enrichment source \citep[e.g.][]{gounelle2012}. Both, winds of Asymptotic Giant Branch stars, which are low-mass stars (M$_* \leq 8$ M$_{\odot}$) at the end of their life spans \citep{wasserburg2006}, as well as winds of Wolf-Rayet stars, which may have been kicked out of a nearby, newly formed star cluster with high-mass stars \citep{gaidos2009}, are possible sources of enriched material. However, all external injection mechanisms have two major downsides: 1) they usually require a specific set of circumstances to occur, such as specific ranges of progenitor masses, close proximity to molecular gas, coincidence of an evolved star or supernova close to the site of new star formation, wind/shock velocities and densities \citep{tachibana2006, ouellette2007, takigawa2008, boss2010, gritschneder2012, boss2019} and 2) they require the enriched (hot) gas to thoroughly mix with the protoplanetary gas within a short timescale \citep{kuffmeier2016}, which seems quite unlikely \citep{seifried2018}.

A local source of enrichment would solve the mixing problem by directly injecting \Alu into the disk immediately before the condensation of dust grains, which typically occurs during the Class I to Class II transition of the young stellar system \citep{dauphas2011}. Cosmic rays (CRs) produced during flare activity in T-Tauri stars have been proposed as a possible mechanism \citep{lee1998}, although their transport through the disk is still poorly constrained \citep{rodgers-lee2017, fraschetti2018, padovani2018}. Enrichment of \Alu acts through highly energetic collisions of parent nuclei, such as \Als and $^{28}$Si, with protons, $\alpha$ particles or more massive CRs. However, the predicted amount of \Alu may still fall short of the measured solar system values \citep{lee1998, gounelle2001}.

In this work, we propose a new mechanism: enrichment via proton irradiation of the surface of protostellar disks during the Class I/II phase by CRs accelerated in the accretion shocks of young protostars. In Section \ref{sec:method} we describe the CR acceleration and attenuation models and the equations solved for the isotope evolution. In Section \ref{sec:res} we present the results of these calculations for models representing different protostellar environments and discuss the broader implications of our results. 

\section{Method}\label{sec:method}
\begin{figure*}[ht!]
    \centering
    \fig{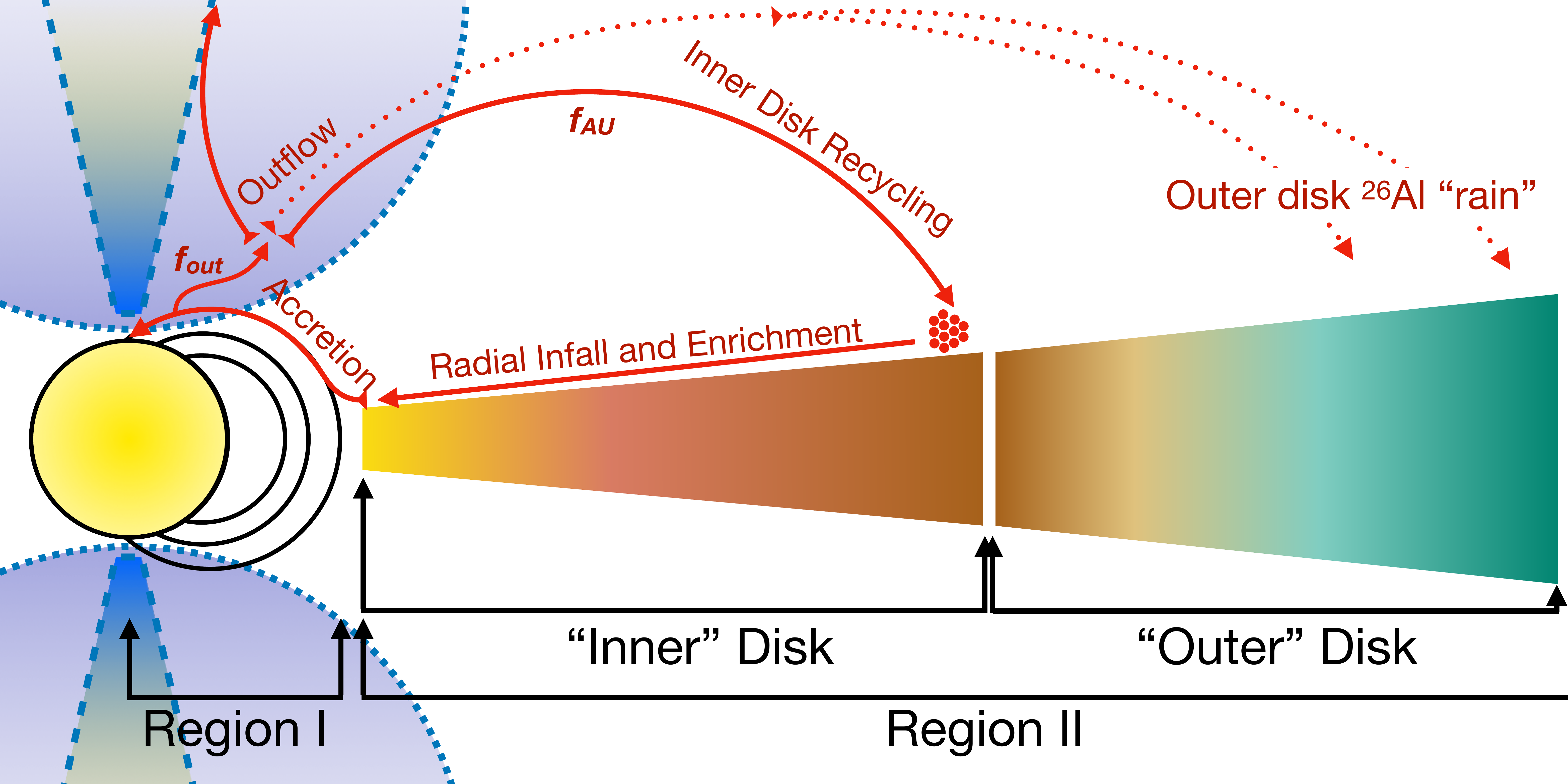}{\textwidth}{}
    \caption{Schematic of the proposed mechanism. We denote the ``gas parcel'' being evolved with the red cluster (see Sections \ref{sec:raddrift} and \ref{sec:isoevol}) . The ``inner'' disk is defined to be the region where $r < 1$ AU. $f_{\rm out}$ is the fraction of the gas swept up by the outflow. $f_{\rm AU}$ is the fraction of the gas in the outflow which falls onto the ``inner'' disk. Some extra fraction of the enriched outflow gas may fall onto the outer disk where the cosmic ray irradiation is weak. We denote the Regions I and II which denote different regimes of cosmic ray transport.}
    \label{fig:schematic}
\end{figure*}

Figure \ref{fig:schematic} shows a schematic of the proposed local enrichment and recycling mechanism. The physical model describes a static disk around a young, accreting central protostar. The gas in the disk drifts radially inwards, where it gets accreted. Cosmic rays are accelerated in the accretion shock. They irradiate the disk and thus, the gas in the inner disk is enriched with \Alu. When the (enriched) material is accreted, some fraction of it, $f_{\rm out}$, is ejected into the protostellar outflow. Finally, a fraction of this outflowing gas rains back down onto the disk, with $f_{\rm AU}$ being the enriched gas fraction, which falls onto the inner disk.

\subsection{Radial Drift Model}\label{sec:raddrift}
We utilize a simple model for inward radial drift. We initialize a gas parcel with $n($\Alu$)/n($\Als$) = 0$ at a radius of 1 AU, where $n$ is the number density of the given species. The parcel is allowed to drift radially inward with a velocity of $v_r = \dot{m}/(2\pi r \Sigma(r))$ until the parcel reaches either the magnetosonic radius or the inner dust rim. $\dot{m}$ is the accretion rate. We use a 1D disk profile where the surface density and temperature are given by
\begin{align}
    \Sigma(r) &= \Sigma_0 \left ( \frac{r}{r_0} \right )^{-p} \\
    T(r) &= T_0 \left ( \frac{r}{r_0} \right )^{-q},
\end{align}
where $\Sigma_0 = 3000$ g cm$^{-2}$, $T_0 = 200$ K, $r_0 = 1$ AU, $p = 1$ and $q = 1/2$. For $p = 1$, the in-fall velocity is constant with radius. The value of $p$ in protostellar disks is not tightly constrained. Circumstellar disks in $\rho$ Ophiuchi show profiles tending towards $p = 1$ \citep{andrews2007} while a more recent survey of dust protoplanetary disks in Lupis found $-0.6 < p < 0.6$ \citep{tazzari2017}. Constraining the disk structure from dust measurements is more complicated due the the fact the dust surface density is likely not a constant factor of the gas surface density \citep{birnstiel2016}. The inner dust rim is defined by where $T_{\rm inner} = 1500$ K, the dust sublimation temperature. During the in-fall, the production rates of \Alu and the isotopic evolution are calculated as described below.

When the parcel reaches the inner disk radius, we assume that the majority of the gas is accreted onto the central object with only a fraction, $f_{\rm out} = 0.1$, of the enriched gas being ejected in the outflow \citep{shu1988, pelletier1992}. We further assume that a fraction, $f_{\rm AU} = 0.1$, of the outflow gas falls back onto the inner disk \citep[see e.g.][]{liffman2005} isotropically. We explore the impact of $f_{\rm AU}$ on the results in \S\ref{sec:depmodel}. After the gas parcel is accreted, another, now slightly \Alu enriched parcel is restarted at $r = 1$  AU, such that the $(i+1)^{\rm th}$ iteration has the following initial condition:
\begin{align}
    f(^{26}{\rm Al})_{i+1} &= f_{\rm AU} f_{\rm out}\left ( \frac{n_i(^{26}{\rm Al})}{n_i(^{27}{\rm Al})} \right ) \\
    n_{i+1}(^{26}{\rm Al}) &= f(^{26}{\rm Al})_{i+1} X({\rm Al}) n_{\rm H} \\
    n_{i+1}(^{27}{\rm Al}) &= (1-f(^{26}{\rm Al})_{i+1}) X_{\rm Al} n_{\rm H}.
\end{align}
Here, $X_{\rm Al}$ is the total aluminium fraction of the gas parcel with respect to hydrogen, %$X = \log{X/H} + 12$, 
and $n_{\rm H}$ is the hydrogen number density. In the following, we denote $f_{\rm tot} = f_{\rm AU} f_{\rm out}$. When the calculation starts, we set $m_* = 0.05$ M$_{\odot}$ and evolve the model until $m_* = 2$ M$_{\odot}$. We consider four different constant accretion rates, $\dot{m} = (10^{-9}, 10^{-8}, 10^{-7}$ and $10^{-6})$ M$_{\odot}$ yr$^{-1}$. We also consider a step-variable tapered accretion history (See Section \ref{sec:burst}).

\subsection{Cosmic Ray Acceleration and Attenuation}

Strong shocks form in accretion columns near the surface of the protostar. Under the assumption of strong ideal hydrodynamic shocks \citep{hartmann2016}, the post-shock temperature and density can be derived from the mass, radius and accretion rate:
\begin{align}
    \label{eq:ts}
    T_s &= \frac{3}{16}\frac{\mu_{\rm I} m_{\rm H}}{k} v_s^2 \\
        &= 1.302\times10^6 \, \left ( \frac{\mu_{\rm I}}{0.6} \right ) \left ( \frac{v_s}{309 {\rm \, km \, s^{-1}}} \right )^2 {\rm \, K} \nonumber ,
\end{align}

\begin{align}
    n_s &= 8.351\times 10^{14} \left ( \frac{\mu_I}{\mu} \right )\left ( \frac{\dot{m}}{10^{-5}  {\rm \, M_{\odot} \, yr^{-1}}} \right ) \times \\ & \left ( \frac{f_{\rm acc}}{0.1} \right )^{-1}
    \left ( \frac{r_*}{2 {\rm \, R_{\odot}}} \right )^{-2} \left ( \frac{v_s}{309 {\rm \, km \, s^{-1}}} \right )^{-1} {\rm \, \, cm^{-3}} \nonumber ,
\end{align}
where $v_s$ is assumed to be the free-fall velocity at the protostellar surface
\begin{align}
    \label{eq:vs}
    v_{\rm s} &= \sqrt{\frac{2Gm_*}{r_*}} \\&= 309 \, \left (\frac{m_*}{0.5 M_{\odot}} \right )^{0.5} \left (\frac{r_*}{2 R_{\odot}} \right )^{-0.5} {\rm \, \, km \, s^{-1}}, \nonumber
\end{align}
where $G$ is the Gravitational constant, $k$ is the Boltzmann constant, $m_H$ is the mass of Hydrogen, $\mu_{\rm I} = 0.6$ is the reduced atomic mass for a fully ionized gas, $\mu$ is an arbitrary reduced atomic mass and $f_{\rm acc}$ is the filling fraction of accretion columns on the surface of the protostar. We calculate the protostellar radius assuming a constant accretion rate history using a standalone protostellar evolution code from \citet{klassen2012}\footnote{\url{https://github.com/mikhailklassen/protostellar_evolution}}, which implements the \citet{offner2009} one-zone model. The accreting mass flows from the inner radius of the disk, which we assume is the maximum of $4r_*$ and the magnetosonic radius,
\begin{equation}
    r_{ms} = \left ( \frac{3B^2 r_*^6}{2\dot{m}(Gm_*)^{1/2}} \right )^{2/7}
\end{equation}
where $B = 10^3$ G is the fiducial protostellar magnetic field. 

We assume the CRs are accelerated via Diffusive Shock Accelerate (DSA), in which protons stochastically cross the shock front numerous times getting impulsive momentum gains at every crossing. In order for DSA to be efficient, the shock must be supersonic and super-Alfv\'{e}nic. The shock is always super-sonic, while it is trans- or sub-Alfv\'{e}nic for high accretion rates and $B_* \geq 10^3$ G. These conditions lead to much lower maximum energies and significant drops in \Alu production, as shown in Section \ref{sec:res}. We adopt the CR acceleration calculation from \citet{gaches2018}, based on \citet{padovani2016}, which produces a non-relativistic spectral slope $\propto E^{-2}$ with a maximum energy of a few GeV. The maximum energy is determined through various time constraints on the acceleration, with energy losses due to interactions with matter, upstream diffusion escape, and magnetic wave dampening dominating. Appendix \ref{sec:appendsa} shows the Alfv\'{e}nic Mach number as a function of mass for different accretion rates. For accretion rates and protostar masses where the shock is super-Alfv\'{e}nic, DSA will be efficient and the magnetic field will not play a significant role in the shock model parameters.

We break up the CR transport into two different regions, following \citet{offner2019}. In Region I, between the shock and the inner disk radius, the CR protons lose energy via Coulomb interactions and pion production \citep{mannheim1994, schlickeiser2002}. We assume the gas is fully ionized with a temperature of $10^4$ K. The column density of the interacting material is calculated by balancing mass accretion with the gas flow at the magnetosonic radius:
\begin{equation}\label{eq:column}
    N_{\rm atten} \equiv \varepsilon \left ( \frac{\dot{m}}{2\pi r_{\rm ms} h \mu m_H} \right ) \left ( \frac{r_{\rm ms}}{v_{\rm s}} \right ),
\end{equation}
where $\varepsilon$ is a free parameter describing how coupled the CRs are to the gas and whether or not the gas flow is intermittent, $h = c_s/\Omega$ is the scale height of the disk, $c_s = \sqrt{\frac{k T}{\mu m_H}}$, and $\Omega = \sqrt{\frac{Gm_*}{r_*^3}}$. We adopt $\varepsilon = 0.1$ for our fiducial model, however, we investigate the effects of lower $\varepsilon$ in Section \ref{sec:res}. The CR flux is further decreased through the transport along magnetic field lines. Following \citet{offner2019}, we adopt the spatial transport factor to be
\begin{align}
    f_M = 0.9 \times \bigg[ 0.5 \left ( \frac{r_*}{r_{\rm ms}} \right )^{-3} &+ 0.3 \left ( \frac{r_*}{r_{\rm ms}} \right )^{-7} \\ &  + 0.1 \left ( \frac{r_*}{r_{\rm ms}} \right )^{-9}  \bigg] \nonumber.
\end{align}
Note that in Region I, the CRs spatial transport is steeper than r$^{-2}$ due to the strong coupling to the magnetic field lines. In Region II, outside the inner disk radius, we assume there is no gas attenuation (gas column densities are typically below 10$^{20}$ cm$^{-2}$). The nature of the transport of CRs from the protostar to the disk is unconstrained. If the magnetic field is turbulent, CRs undergo a more diffusive transport, $r^{-1}$. If the magnetic field is weaker, the CRs will free-stream, $\propto r^{-2}$ \citep{schlickeiser2002}. We assume the CRs have a transport as $r^{-2}$ to be conservative (See \S\ref{sec:depmodel}).

Therefore, the CR spectrum at the magnetosonic radius, $f(E, r_{\rm ms})$, is
\begin{equation}
f(E, r_{\rm ms}) = f(E, r_*)g(N_{\rm atten})\times f_M,
\end{equation}
where $f(E, r_*)$ is the CR spectrum at the protostellar surface and $g(N_{\rm atten})$ accounts for the attenuation by column density $N_{\rm atten}$. The CR spectrum at the disk within Region II is
\begin{equation}
    f(E, r) = f(E, r_{\rm ms}) \times \left ( \frac{r_{\rm ms}}{r} \right )^2.
\end{equation}
We only consider the interactions of CRs with the gas at the surface of the protostellar disk and therefore do not model the propagation and attenuation into the disk, which can be complicated due to the importance of secondary and tertiary processes \citep{padovani2018}.

\subsection{Nuclear Cross Sections and Production Rate}
\begin{figure}
    \centering
    \includegraphics[width=0.5\textwidth]{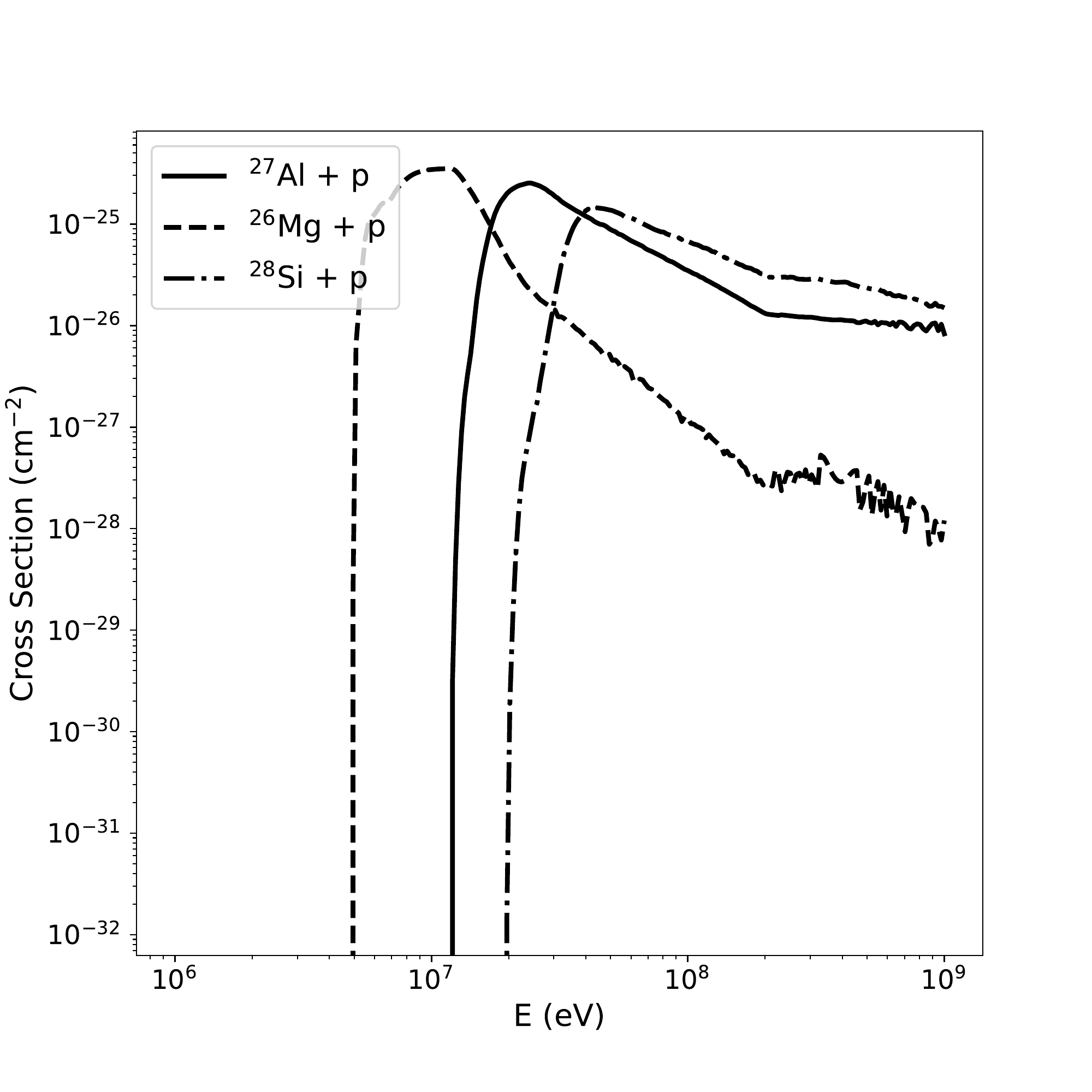}
    \caption{Cross sections as a function of energy for \Alu production via \Als, $^{26}$Mg and $^{28}$Si calculated by the {\sc Talys} code.}
    \label{fig:xsec}
\end{figure}

We calculate the production rate, $\zeta$, of \Alu at the surface of the protostellar disk via some process, $i$, due to interactions with CRs, such that
\begin{equation}
    \zeta(i) = \int_{E_{th}}^{\infty} f(E) \sigma_i(E) dE,
\end{equation}
where $f(E)$ is the CR spectrum and $\sigma_i(E)$ is the interaction cross section for process, $i$. We take into account three different processes:
\begin{itemize}
    \item $^{27}{\rm Al} + p \rightarrow ^{26}{\rm Al}$ + pn,
    \item $^{26}{\rm Mg} + p \rightarrow ^{26}{\rm Al}$ + n,
    \item $^{28}{\rm Si} + p \rightarrow ^{26}{\rm Al}$ + $^{3}$He
\end{itemize}
We only account for interactions with protons and not heavier species such as $\alpha$-particles. Figure \ref{fig:xsec} shows the nuclear cross sections calculated by the {\sc Talys} code \citep{koning2012}. 

\subsection{Equations for Isotopic Evolution}\label{sec:isoevol}

We evolve the isotopic abundances of \Alu and \Als using brute-force first-order integration of the following system of equations: 
\begin{align}
    \frac{dn(^{26}{\rm Al})}{dt} &= n(^{27}{\rm Al})\zeta(^{27}{\rm Al})  + n(^{26}{\rm Mg})\zeta(^{26}{\rm Mg}) + \nonumber \\
    & n(^{28}{\rm Si})\zeta(^{28}{\rm Si}) - \frac{n(^{26}{\rm Al})}{\tau_{^{26}{\rm Al}}}  \\
    \frac{dn(^{27}{\rm Al})}{dt} &= -n(^{27}{\rm Al})\zeta(^{27} {\rm Al}) \\
    \frac{dn(^{26}{\rm Mg})}{dt} &= -n(^{26}{\rm Mg})\zeta(^{26} {\rm Mg}) +  \frac{n(^{26}{\rm Al})}{\tau_{^{26}{\rm Al}}} \label{eq:26Mg} \\
    \frac{dn(^{28}{\rm Si})}{dt} &= -n(^{28}{\rm Si})\zeta(^{28} {\rm Si}).
\end{align}
The second term on the right hand side of Equation \ref{eq:26Mg} comes from the decay of \Alu into $^{26}$Mg. We adopt a gas density at the surface of the disk of $n_g = 10^5$ cm$^{-3}$. We use solar abundances, given as $X(i) = \log \frac{n(i)}{n(H)} + 12$, of aluminium $X({\rm Al})$ = 6.43, magnesium $X({\rm Mg})$ = 7.53, and silicate $X({\rm Si})$ = 7.51 and assume an initial $^{26}$Mg isotope fraction of 0.11 \citep{asplund2009}. The initial condition $n_0(^{26}{\rm Al})/n_0(^{27}{\rm Al}) = 0$ is imposed as a ``worst-case'' scenario, although it could start significantly higher due to some pre-enrichment during the earlier stages of protostar formation. Our models do not take into account dust grain morphology or vertical settling. Furthermore, our models assume there is a definitive disk surface, which may not be the case, where the core transitions into a rotating disk or the disk has a fluffy surface. However, this would not impact the CR population as the column density between the magneto-sonic radius and the regions of the disk we consider are below 10$^{20}$ cm$^{-2}$.

\section{Results and Discussion}\label{sec:res}
We consider four different models (see Table~\ref{tab:dmodels}). The fiducial model uses $n_g = 10^5$ cm$^{-3}$, $\varepsilon = 0.1$, $f_{\rm AU} = 0.1$, f$_{\rm out} = 0.1$ (therefore $f_{\rm tot} = 0.01$) and $B_* = 10^3$ G. These parameters are chosen to best represent a protostellar disk towards the end of the Class I phase with conservative estimates $B_*$ and $\varepsilon$. We note that there is little difference in the main results for $n_g = 10^4 - 10^5$ cm$^{-3}$. We consider three other models which vary major parameters in this work. Models 2-4 in Table \ref{tab:dmodels} have all the same parameters as the fiducial model except for the difference in the description column. 

\begin{deluxetable}{cc}[ht!]
\label{tab:dmodels}
\tablecolumns{2}
\tablecaption{ Models}{}
\tablehead{\colhead{Model \#} & \colhead{Description} }
\startdata
1 & Fiducial \\
2 & B$_* = 10^1$ G \\
3 & $\varepsilon = 10^{-3}$ \\
4 & $f_{\rm tot} = 0.25$ \\
\enddata
\tablecomments{Models used in the paper.}
\end{deluxetable}

Figure \ref{fig:evol_ratio} shows the \Alu/\Als ratio and the protostellar radius as a function of the protostellar mass for different accretion rates. Low levels of accretion are able to generate sufficient \Alu to match meteor enrichment ratios \citep{jacobsen2008}. Large accretion rates are ineffective at producing \Alu due the enhanced shock column densities and protostellar radii. We find that \Alu/\Als correlates with the radius for a given accretion rate. Abrupt changes in $r_*(m_*)$ produce similar behavior in the enrichment ratio. Low accretion rates, $\dot{m} < 10^{-7}$ M$_{\odot}$ yr$^{-1}$, produce significant amounts of \Alu in the inner disk due the minimal column density in the accretion region and  smaller protostellar radii. It is worth noting that, in general, low-mass protostars undergo these levels of accretion during their transition from Class I to Class II objects \citep{hartmann2016}, a time period during which dust begins to condense and planet formation commences \citep{tobin2020}.  

\subsection{Dependence on Model Parameters}\label{sec:depmodel}
\begin{figure*}[t!]
    \fig{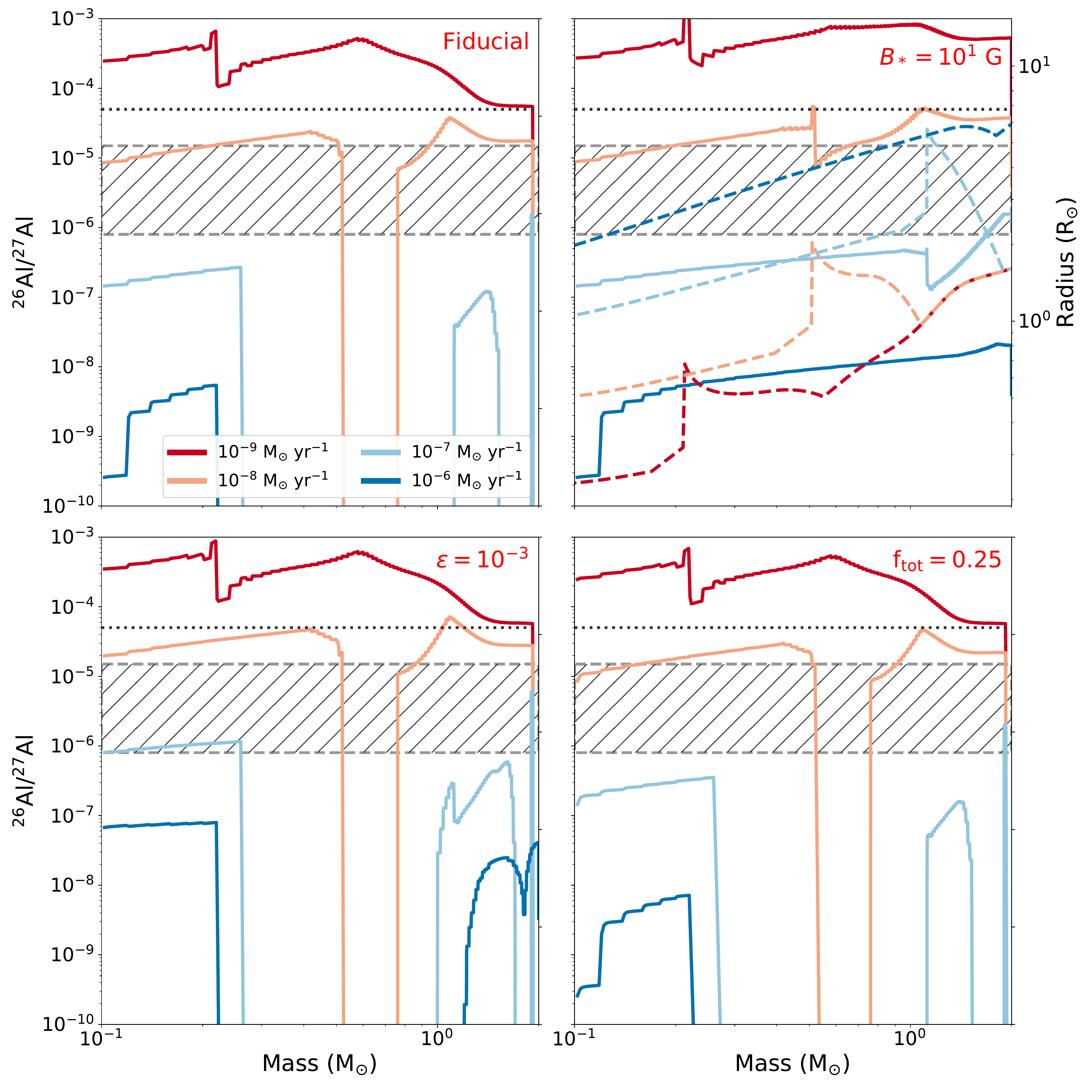}{\textwidth}{}
    \caption{\Alu/\Als as a function of protostellar mass for different constant accretion rates, denoted by line color. The panels correspond to Models 1, 2, 3 and 4 in the top left, top right, bottom left and bottom right, respectively, from Table \ref{tab:dmodels}. The colored dashed lines in the top right panel show the protostellar radius (see left y-axis) as a function of mass for the accretion rate denoted by line color. The grey dotted line highlights the canonical measured meteorite ratio \Alu/\Als $= 5 \times 10^{-5}$. The grey hatched region shows the bounds defined in \citet{goswami2005} for measurements of chondritic meteorites and differentiated meteors.} 
    \label{fig:evol_ratio}
\end{figure*}

The top panels of Figure~\ref{fig:evol_ratio} show the impact of different protostellar magnetic field strengths. The magnetic field at the surface of protostars is still unknown. While our fiducial model assumes $B_* = 10^3$ G to match T-Tauri stars \citep{johns-krull2007}, the magnetic field may be substantially lower during earlier phases of star formation. We find that decreasing the protostellar magnetic field to 10 G leads to less than an order of magnitude increase in \Alu/\Als for $\dot{m} \leq 10^{-8}$ M$_{\odot}$ yr$^{-1}$. For higher accretion rates, the lower magnetic field results in an increase by many orders of magnitude. This is due to the smaller magnetosonic radius and more efficient acceleration of CRs at these magnetic field strengths \citep[see][]{gaches2018}. Thus, the protostellar magnetic field and, in particular, radius are two of the most crucial factors in the production of \Alu. The top right panel of Figure \ref{fig:evol_ratio} shows that a rapid increase in the protostellar radius (and thus magnetosonic radius) as a function of mass suddenly removes all enrichment from the disk. Since $f_M$ is a steep function of radius, the increase in magneticsonic radius greatly damps the amount of CRs impinging the disk. Furthermore, an increased radius leads to weaker accretion shocks.

We also include two models investigating the importance of the CR-gas coupling and \Alu recycling efficiencies. Reducing the coupling parameter, $\varepsilon = 10^{-3}$, provides only a factor two increase in \Alu/\Als at low accretion rates. The proton range around MeV energies is superlinear as a function of energy, and therefore decreasing the column by two orders of magnitude produces less than an order of magnitude increase in proton flux. For high accretion rates, this reduction leads to an increase of several orders of magnitude for the protostellar masses (radii), which allow CRs to reach the inner disk. The degree of mixing only marginally impacts the results. In fact, as a test, a model was run with $f_{\rm tot} = 0$ (i.e. no recycling), which resulted in only a marginal reduction in \Alu/\Als. The enrichment can occur on the accretion timescale, $t_{\rm acc} \propto r/v$, which may be substantially faster than the formation timescale of the protostar. The enrichment, thus, would depend on the accretion history on these timescales. We have found that the the main conclusion of this work is not greatly sensitive to $f_{\rm AU}$. However, \Alu leaving the protostellar system through the outflows necessarily depends on the accretion history.

We have assumed CRs transport is ballistic in Region II. However, their transport may be more diffusive. We ran a test model with the fiducial parameters but with diffusive transport, $r^{-1}$, in Region II and found $^{26}$Al/$^{27}$Al to increase by an order of magnitude for all accretion rates which allowed for efficient DSA.

We assume a power-law gas surface density profile, $\Sigma \propto r^{-p}$ with $p = 1$. However, as mentioned above, the value of $p$ is not well constrained. Shallower profiles lead to a slightly faster accretion time, but for the accretion rates which we find are important for \Alu production, the difference is marginal compared to overall timescale. A shallower profile will lead to a decrease in the amount of \Alu produced in the disk. Overall, the density distribution index is not a dominant source of uncertainty, compared to the dependence on the protostellar radius and magnetic field.

\subsection{Step-Tapered Accretion}\label{sec:burst}
\begin{figure*}[t!]
    \fig{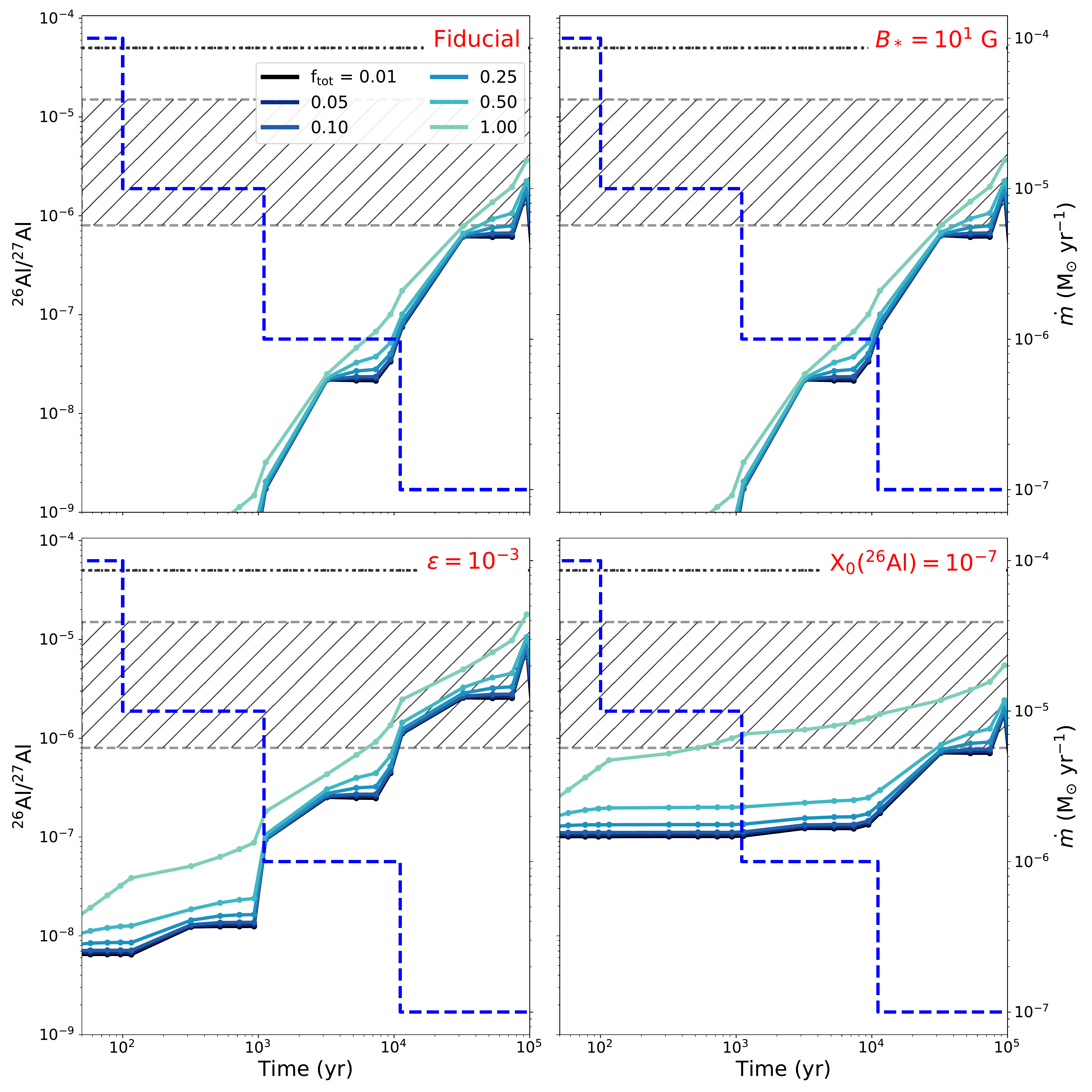}{\textwidth}{}
    \caption{Evolution of \Alu/\Als and accretion rate for the step-tapered accretion model. The panels correspond to Models 1, 2, 3 and 4 in the top left, top right, bottom left and bottom right, respectively, from Table \ref{tab:dmodels}. The line color denotes $f_{\rm tot} = f_{\rm out}f_{\rm AU}$ as designated in the legend. The blue dashed line shows the evolution of the accretion rate. The grey dotted line highlights the canonical measured meteorite ratio \Alu/\Als $= 5 \times 10^{-5}$. The grey hatched region shows the bounds defined in \citet{goswami2005} for measurements of chondritic meteorites and differentiated meteors.} 
    \label{fig:stepview}
\end{figure*}

Protostellar accretion rates likely start high and decline over time as the natal core and disk mass are depleted \citep{fischer2017}, with Class II protostars having accretion rates $\dot{m} < 10^{-9}$ M$_{\odot}$ yr$^{-1}$ \citep{hartmann2016}. To investigate the impact of accretion changes on our results,  we adopt a simple step function history where the accretion rates, $\dot{m} = (10^{-4}, 10^{-5}, 10^{-6}, 10^{-7})$  M$_{\odot}$ yr$^{-1}$ are pulsed for $\Delta t = (10^2, 10^3, 10^4,$ and $10^5)$ years, respectively, for a total of 0.04 M$_{\odot}$ of accretion over 0.11 Myr. The initial protostellar mass in this model is $m_* = 0.9$ M$_{\odot}$, to imitate the proto-Sun. We evolve the step-tapered history using the described radial drift model. Figure \ref{fig:stepview} shows the temporal evolution of \Alu/\Als and the accretion rate (dashed line). We find that \Alu is only substantially produced when the accretion rates are low, similar to the above analysis. While all models exhibit \Alu/\Als similar to the Chondrite measured range \citep{goswami2005}, only the model with low CR-gas coupling in the inner disk is able to produce \Alu/\Als ratios near meteor measurements. 

We include a model (bottom right) where the initial \Alu abundance, $n_0({\rm ^{26}Al})/n_0({\rm ^{27}Al}) = 10^{-7}$, is a fraction of the Galactic mean value to represent primordial enrichment of the natal molecular gas. The external contamination keeps \Alu/\Als at a near constant value until $\dot{m} < 10^{-6}$ M$_{\odot}$ yr$^{-1}$ for $f_{\rm tot} \neq 1$. The final ratio is not sensitive to small amounts of primordial \Alu abundance enrichment.

\subsection{Discussion}\label{sec:conc}
We have shown that CRs accelerated in the accretion shocks of protostars may provide significant enrichment of aluminum-26 in the inner regions of protostellar disks. Calcium-aluminum-rich inclusions (CAIs) likely condense rapidly after the end of the protostellar Class~I phase in the beginning of the Class II phase, as the inner disk cools below 1500 K \citep{adams2010, dauphas2011}. \Alu is thought to be an important source of heating during planet formation, allowing the generation of differentiated planetesimals with metal cores \citep{gail2014}. Our models show that a proto-Sun can accelerate enough CRs to produce substantial amounts of \Alu before chondrule formation. We find that only if the gas flow between the disk and protostar is intermittent, spatially or temporally, or the CRs are not fully coupled to the magnetic field, will \Alu/\Als approach meteorite measurements. Accretion rates $\dot{m} \leq 10^{-8}$ M$_{\odot}$ yr$^{-1}$ rapidly lead to \Alu/\Als $> 5 \times 10^{-5}$. Figures~\ref{fig:evol_ratio} and \ref{fig:stepview} show that the amount of enrichment is not very sensitive to the protostellar magnetic field or the degree to which \Alu mixes with the non-enriched gas after accretion, except for high-accretion periods. The proposed mechanism would lead to an initially heterogeneous distribution of \Alu (it might homogenize later during turbulent mixing of the disk), proposed by some (see above), leading to an incomplete chemical homogenisation. Isotope compositions of early solar system matter are broadly homogeneous, although there is a clearly discernable difference between the inner and outer protoplanetary disk, as inferrend from nucleosynthetic isotope variations in chondrites \cite[see][]{warren2011}.

External events are enticing and plausible mechanisms for enriching the early solar system with \Alu. However, they often necessitate specific circumstances, suggesting that systems with enrichment similar to that of the Solar System are rare. This can be alleviated by the mechanism proposed here. An internal source of CRs provides enough particle irradiation to prevent substantial decay of the \Alu density. Counter-intuitively, the enrichment is enhanced when the accretion rates are low. This is due to the strong impact that the attenuating gas within the accreting gas has on the resulting \Alu production rate. Higher accretion rates also enhance the protostellar radius, thereby increasing the magnetosonic radius. However, low levels of accretion have a small attenuating column and a smaller radius and are therefore not strongly damped by the time they reach the inner disk. Protostars likely undergo periods of low accretion during the transition between Class I and Class II phases. Figure \ref{fig:evol_ratio} shows the proposed mechanism is general for a wide range of low-mass stars, from M-dwarfs to solar-type stars, where the majority of exoplanetary systems have been discovered. As such, {\it  CRs accelerated by accretion onto protostars may provide a general pathway for \Alu enrichment in many planetary systems.}

Some proposed enriched mechanisms, such as Wolf-Rayet star and Asymptotic Giant Branch star winds, may also provide \Alu during the Class II phase. However, these two sources require strict constraints on the physical setup between the protoplanetary disk and wind location. Our models show that these constraints may be alleviated by requiring much less \Alu transported to the disk from the outside as long as the protostar undergoes a phase of low-level accretion (which all of them certainly do). However, external enrichment mechanisms may not be necessary, as actively accreting protostars can produce enough \Alu in under a Myr to seed protoplanetary disks with this important radioactive isotope. Our models are one-dimensional and make assumptions, such as the assumption of a static disk. In reality, the disk structure may be non-axisymmetric and evolves over time. Furthermore, the accretion column may be complicated, leading to a dependency of $\epsilon$ on the protostellar mass or accretion rate. However, the simple models presented here show the potential importance of CRs accelerated by accretion shocks.

Protostars are rarely isolated, typically forming in clusters with a wide range of sizes, from a few protostars to thousands. In this work, we have only considered the enriched material falling back onto the protostellar disk. However, a fraction of enriched accreted gas, $f_{\rm out}(1-f_{\rm AU}) \approx f_{\rm out}$, would escape the protostellar system and leak into the rest of the molecular cloud. As such, protostellar systems in clusters may have higher initial abundances of \Alu due to nearby protostellar jets. A calculation of this form of cluster-enrichment is beyond the scope of the paper, however, it is an intriguing possibility similar to the hierarchical star formation mechanism posited by \citet{fujimoto2018}. \Alu enrichment via MeV CRs accelerated in protostellar accretion shocks thus may be an important source of \Alu (and other SLRs) during the star and planet formation process as well as in our own Solar System.

\acknowledgments
The work presented here was inspired by fruitful discussions with Megan Reiter after the CRC 956 Colloquium at the Universit\"{a}t zu K\"{o}ln on 18 November 2019, and we therefore thank the German Science Foundation for support via SFB 956, project C5. We thank the anonymous referee for their useful comments improving this work. This work was funded by the ERC starting grant No. 679852 ‘RADFEEDBACK’. S.S.R.O. acknowledges funding from NSF Career grant AST-1650486 and NASA ATP grant 80NSSC20K0507. This is University of Texas Center for Planetary Systems Habitability Contribution \#0008

\appendix
\section{Validity of DSA Conditions}\label{sec:appendsa}
Diffusive shock acceleration (DSA) is efficient only when the shock is super-sonic and super-Alfv\'{e}nic. The shock model described in Section \ref{sec:method} assumes a strong ideal shock, and therefore, the flow is always supersonic. However, there is no guarantee for the Alfv\'{e}n Mach number. Both, the radius and magnetic field, where neither is well constrained by observations, significantly impact the Alfv\'{e}n Mach number of the shock. The Alfv\'{e}n speed is strongly dependent on the magnetic field and radius and weakly dependent on the mass: $v_A \propto B_*/\sqrt{n_s} \propto B_* r_*^{3/4} m_*^{1/4}$, since $n_s \propto r_*^{-3/2} m_*^{-1/2}$. Furthermore, the shock velocity  $v_{\rm s} \propto m_*^{1/2} r_*^{-1/2}$. As such, $\mathcal{M}_A = v_{\rm s}/v_A \propto m_*^{1/4} B_*^{-1} r_*^{-5/4}$.

\begin{figure}
    \centering
    \fig{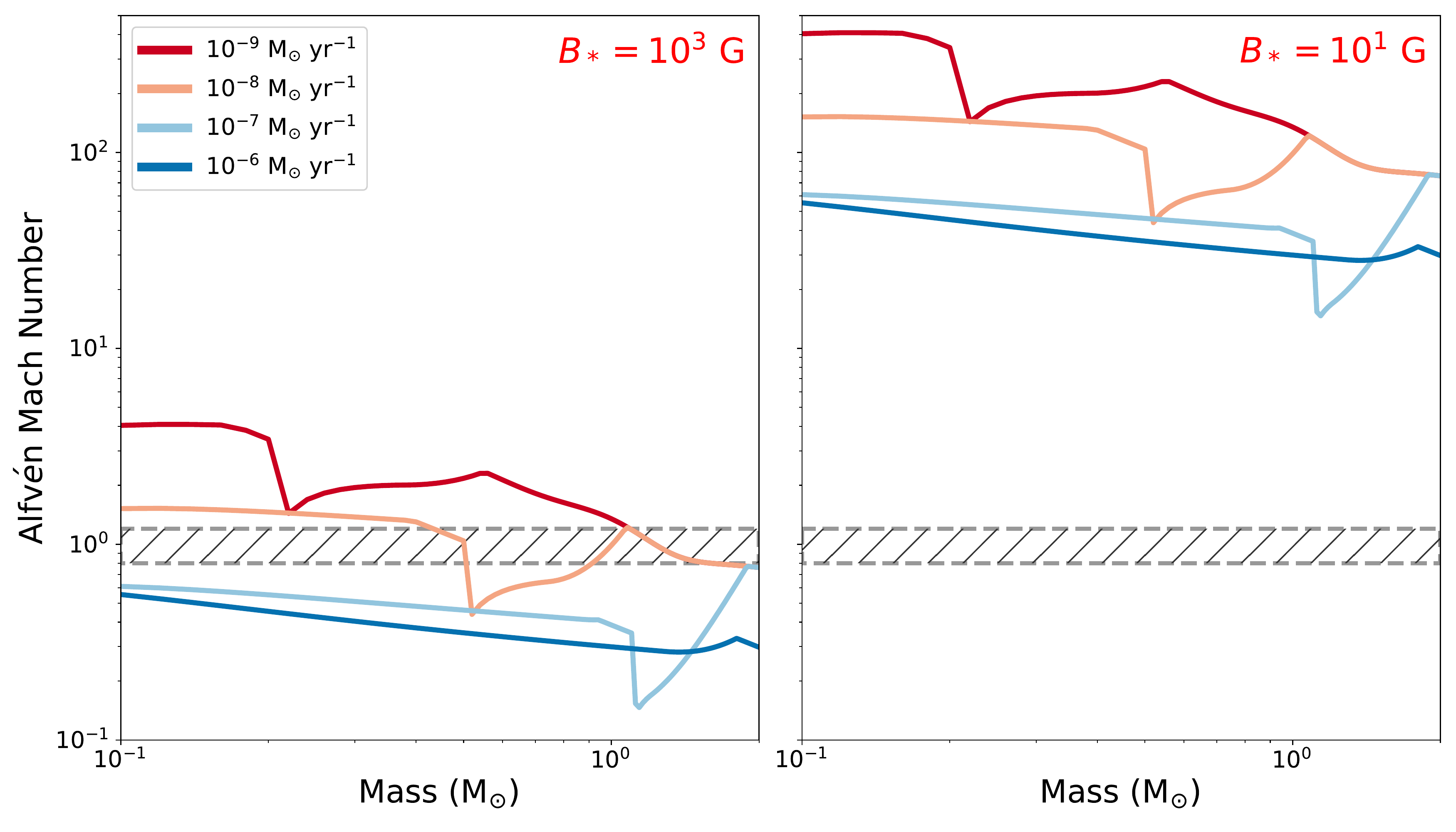}{\textwidth}{}
    \caption{Alfv\'{e}n Mach number as a function of mass for the accretion rates considered in this paper. The grey band indicates the trans-Alfv\'{e}nic region. Left: Protostellar magnetic field, $B_* = 10^3$ G. Right: Protostellar magnetic field, $B_* = 10$ G.}
    \label{fig:alfv_mach}
\end{figure}

Figure \ref{fig:alfv_mach} shows the Alfv\'{e}n Mach number as a function of mass for the accretion rates considered in this paper. We find that high accretion rates produce sub-Alfv\'{e}nic shocks due to the large protostellar radii. For a protostellar magnetic field of $B_* = 10^3$ G, and higher accretion rates, $\dot{m} > 10^{-7}$ M$_{\odot}$ yr$^{-1}$ the shock is sub-Alfv\'{e}nic at all masses. However, for a lower magnetic field, $B_* = 10$ G, the shock is highly super-Alfv\'{e}nic for all accretion rates and masses. Therefore, the protostellar magnetic field is a crucial parameter to constrain. We interpret the regions in Figure \ref{fig:evol_ratio} where \Alu/\Als $\rightarrow 0$ as a failure to accelerate protons out of the thermal tail due to trans- or sub-Alfv\'{e}nic shocks. Figures \ref{fig:evol_ratio} and \ref{fig:alfv_mach} show this is alleviated with lower protostellar magnetic fields.

\software{{\sc Scipy} \citep{scipy2001},
          {\sc Numpy} \citep{numpy2011} ,
          {\sc Matplotlib} \citep{matplotlib2007},
          {\sc JupyterLab}}
          
\bibliography{lib}
\bibliographystyle{aasjournal}



\end{document}